\documentclass{article}
\usepackage{amsmath}
\usepackage{graphicx}
\usepackage{amsfonts}
\usepackage{amssymb}
\usepackage{geometry}
\usepackage{float}
\usepackage{bm}
\usepackage{color}
\usepackage{cite}

\begin{document}

\title{Dunkl-Klein-Gordon equation in three-dimensions: \\The Klein-Gordon
oscillator and Coulomb Potential}
\author{B. Hamil\thanks{%
hamilbilel@gmail.com} \\
D\'{e}partement de TC de SNV, Universit\'{e} Hassiba Benbouali, Chlef,
Algeria. \and B. C. L\"{u}tf\"{u}o\u{g}lu\thanks{%
bclutfuoglu@akdeniz.edu.tr} \\
Department of Physics, Akdeniz University, Campus 07058, Antalya, Turkey, \\
Department of Physics, University of Hradec Kr\'{a}lov\'{e}, \\
Rokitansk\'{e}ho 62, 500 03 Hradec Kr\'{a}lov\'{e}, Czechia.}
\date{}
\maketitle

\begin{abstract}
Recent studies show that deformations in quantum mechanics are inevitable. In this contribution, we consider a relativistic quantum mechanical differential equation in the presence of Dunkl operator-based deformation and we investigate solutions for two important problems in three-dimensional spatial space. To this end, after introducing the Dunkl quantum mechanics,  we examine the Dunkl-Klein-Gordon oscillator solutions with the Cartesian and spherical coordinates. In both coordinate systems,  we find that the differential equations are separable and their eigenfunctions can be given in terms of the associate Laguerre and Jacobi polynomials. We observe how the Dunkl formalism is affecting the eigenvalues as well as the eigenfunctions.  As a second problem, we examine the Dunkl-Klein-Gordon equation with the Coulomb potential. We obtain the eigenvalue, their corresponding eigenfunctions, and the Dunkl-fine structure terms.
\end{abstract}

\section{Introduction}
Nearly a century after the establishment of quantum mechanics, we observe an increasing number of studies on the deformations of the classical formalism of Heisenberg algebra \cite{Mangano,Hinrichsen, Nozari, Nozari1, Nouicer}. All considered deformations rely on different physical motivations, for example, a minimal observable length, maximally observable momentum and length, etc...\cite{Pedram2012, Perivolaropoulos, chung, ek7}. In the last decade, a new deformation is under investigation in quantum mechanics by substituting the Dunkl derivative with the standard spatial one. In fact, the history of the Dunkl operator goes back to the last century. In 1950, Wigner suggested a new method to perform the quantization of bosons. According to him, some constants later called Wigner parameters, and parity operators could be employed in the ordinary commutator of position and momentum. Moreover, such deformation was keeping the equation of motion invariant \cite{Wigner1950}. One year later,  Yang applied the deformed algebra to solve one dimensional quantum harmonic oscillator problem \cite{Yang1951}. In 1989, Charles Dunkl introduced a new operator by combining the usual differential and parity operators with Wigner parameter \cite{Dunkl1, Dunkl2}. Dunkl operator have been employed in various sectors of mathematics \cite{Heckman, Bie}.  Since Dunkl derivatives are related to the Bannai-Ito \cite{Ito} and Schwinger-Dunkl \cite{Genest1, Genest2} algebras, it provides additional information in different areas of physics. For example,  generalization of the Cologero-Moser-Sutherland models \cite{CMS1,CMS2}, isotropic oscillator in two and three dimension  \cite{Genest1, Genest2, Genest3}, Coulomb potential problem on the plane \cite{Genest4}, Dirac  \cite{Mota1}, Klein-Gordon  \cite{Mota2},  pseudo-harmonic \cite{Mota3}, and  Duffin-Kemmer-Petiau  \cite{Merad} oscillator problems are examined comprehensively. Besides, Chung et al. discussed one dimensional Dunkl-quantum mechanics with the simplest example a confined particle in  one-dimensional box in \cite{CHU1, CHU2}. Kim et al. extended electrostatics to Dunkl formalism in \cite{KIM}, while Ghazouani et al. solved Dunkl-Coulomb problem in three dimensions \cite{Sami}. Mota et al. derived the Landau energy levels of the relativistic Dunkl oscillator in \cite{Mota4}. Very recently, statistical mechanics are also revisited with Dunkl deformation \cite{sta1, sta2}.

In this contribution we work in three spatial dimensions, at first, we examine the Dunkl-Klein-Gordon oscillator within Cartesian coordinates, then in spherical coordinates. Then, we consider the Coulomb potential and derive a solution out of the Dunkl-Klein-Gordon equation. To our best knowledge, Mota et al. studied a similar problem in two dimensions with the polar coordinates \cite{Mota2}, and nobody handled this problem in three-dimensional geometry. Since we live in three spatial dimensions, we think our findings will be very important.

We organize the manuscript as follows: In the next section, we introduce the three Dunkl algebra and quantum mechanics. Then, in section \ref{sec:3}, we solve Dunkl-Klein-Gordon oscillator in three dimension in Cartesian and spherical coordinates, respectively. Next, in section \ref{sec:4}, we study Coulomb potential in Klein-Gordon equation within Dunkl formalism. Finally, we conclude the manuscript in section \ref{sec:5}.

\section{Dunkl-Quantum Mechanics}

\bigskip In the Dunkl quantum mechanics the ordinary partial derivatives are
substituted with the Dunkl derivatives 
\begin{equation}
D_{j}=\frac{\partial }{\partial x_{j}}+\frac{\mu _{j}}{x_{j}}\left(
1-R_{j}\right) .
\end{equation}%
Here,  Wigner constants, $\mu _{j}$ are positive real numbers. $R_{j}$ are the reflection operators which satisfy the following action \cite{Genest1,Genest2}:
\begin{equation}
R_{j}f\left( x_{j}\right) =f\left( -x_{j}\right) ;\quad \quad R_{i}R_{j}=R_jR_i;\quad\quad R_{j}x_i=-\delta_{ij}x_i R_j ; \quad \quad R_{j}\frac{\partial }{\partial x_{j}}=-\delta_{ij}\frac{\partial }{\partial x_{i}}R_{j} . \quad (\text{no summation})
\end{equation}%
Therefore, the Dunkl operators obey the following algebra: 
\begin{equation}
 R_jD_j=-D_jR_j;\quad\quad
\left[ D_{i},D_{j}\right] =0; \quad\quad \left[ x_{i},D_{j}\right] =\delta _{ij}\left( 1+2\mu _{_{\delta_{ij}}}R_{_{\delta_{ij}}}\right). \quad (\text{no summation})
\end{equation}%
Before proceeding to the next section, it would be useful to highlight some other features of the Dunkl operator.
\begin{itemize}
\item It is a linear operator:
\begin{equation}
D_{j}\left[ af\left( x_{j}\right) +bg\left( x_{j}\right) \right]
=aD_{j}f\left( x_{j}\right) +bD_{j}g\left( x_{j}\right) ,
\end{equation}

\item For any $f\left( x_{j}\right) ,$ $g\left( x_{j}\right) $ the Dunkl
derivative satisfies the general Leibniz rule \cite{Micho}:
\begin{equation}
D_{j}\left( f\left( x_{j}\right) g\left( x_{j}\right) \right) =\left(
D_{j}f\left( x_{j}\right) \right) g\left( x_{j}\right) +f\left( x_{j}\right)
\left( D_{j}g\left( x_{j}\right) \right) -\frac{\mu _{j}}{x_{j}}\left[
\left( 1-R_{j}\right) f\left( x_{j}\right) \right] \left[ \left(
1-R_{j}\right) g\left( x_{j}\right) \right] .
\end{equation}%
If one of these functions is even, ($f\left( x_{j}\right) $ or $g\left(
x_{j}\right) $), then we obtain the usual Leibniz formula.
\begin{equation}
D_{j}\left( f\left( x_{j}\right) g\left( x_{j}\right) \right) =\left(
D_{j}f\left( x_{j}\right) \right) g\left( x_{j}\right) +f\left( x_{j}\right)
\left( D_{j}g\left( x_{j}\right) \right) .
\end{equation}
\item In the presence of Dunkl derivatives, the ordinary definition of the scalar product must be modified with \cite{CHU1}:
\begin{equation}
\left \langle f\right. \left \vert g\right \rangle =\int dxg^{\ast }\left(
x\right) f\left( x\right) \left \vert x\right \vert ^{2\mu }dx,  \label{scalar}
\end{equation}%
\item The expectation value of an operator $\mathcal{O}$ with respect to the
state, $\psi$, can be defined by%
\begin{equation}
\left \langle \psi \right \vert \mathcal{O}\left \vert \psi \right \rangle
=\int dx\psi ^{\ast }\left( x\right) \mathcal{O}\psi \left( x\right)
\left \vert x\right \vert ^{2\mu }dx.
\end{equation}

\item The square of the Dunkl derivative can be expressed in the form of
\begin{equation}
D_{i}^{2}=\frac{\partial ^{2}}{\partial x_{i}^{2}}-\frac{\mu _{i}}{x_{i}^{2}}%
\left( 1-R_{i}\right) +\frac{2\mu _{i}}{x_{i}}\frac{\partial }{\partial x_{i}%
}.
\end{equation}
\end{itemize}

\section{Dunkl-Klein-Gordon oscillator in three dimensions}\label{sec:3}

In this section, we solve the Dunkl-Klein-Gordon oscillator in three dimensions. To this end, we employ 
the Dunkl momentum, $\frac{1}{i}D_{j}$, instead of the ordinary momentum operator, $p_{j}$, in natural units. In that case, the stationary Klein-Gordon oscillator equation reads
\begin{equation}
\left \{ E^{2}-\left( \frac{1}{i}D_{j}+im\omega x_{j}\right) \left( \frac{1}{i%
}D_{j}-im\omega x_{j}\right) -m^{2}\right \} \psi =0;\text{ \ with }j=%
\overline{1;3},
\end{equation}%
where $m$ and $\omega $ indicate the rest mass and  oscillator frequency, respectively. When we follow the Dunkl algebra which is summarized in the previous section,  we get the Dunkl-Klein-Gordon oscillator equation in Cartesian coordinates.
\begin{equation}
\left \{ -D_{1}^{2}-D_{2}^{2}-D_{3}^{2}-2m\omega \left( \mu _{1}R_{1}+\mu_{2}R_{2}+\mu _{3}R_{3}+\frac{3}{2}\right) +m^{2}\omega ^{2}\left(
x_{1}^{2}+x_{2}^{2}+x_{3}^{2}\right) \right \} \psi =\left(
E^{2}-m^{2}\right) \psi .  \label{A}
\end{equation}
This equation is manifestly separable in Cartesian and spherical coordinates even in the presence of the reflection operators.

\subsection{Solution in Cartesian coordinates}
By using the Cartesian coordinates, we can obtain three one-dimensional Dunkl-Klein-Gordon oscillators out of Eq. \eqref{A}. To this end, we define
\begin{eqnarray}
E^{2}-m^{2}&=&\mathcal{E}_{1}+\mathcal{E}_{2}+\mathcal{E}_{3}, \label{E} \\   
\psi&=& \psi\left( x_{1}\right) \psi\left( x_{2}\right)
\psi\left( x_{3}\right), \\
\mathcal{H}&=&\mathcal{H}_{1}+\mathcal{H}_{2}+\mathcal{H}_{3}, 
\end{eqnarray}
where
\begin{equation}
\mathcal{H}_{j}=-D_{j}^{2}-m\omega \left( 1+2\mu _{j}R_{j}\right)
+m^{2}\omega ^{2}x_{j}^{2}, \quad j=1,2,3.
\end{equation}%
Then, we could be able to express the Dunkl-Klein-Gordon oscillator as it is one dimension with the given energy eigenvalue constraint.
\begin{equation}
\left \{ \frac{\partial ^{2}}{\partial x_{j}^{2}}+\frac{2\mu _{j}}{x_{j}}%
\frac{\partial }{\partial x_{j}}-\frac{\mu _{i}}{x_{j}^{2}}\left(
1-R_{j}\right) -m^{2}\omega ^{2}x_{j}^{2}+m\omega \left( 1+2\mu
_{j}R_{j}\right) -\mathcal{E}_{j}\right \} \psi \left( x_{j}\right) =0.
\label{B}
\end{equation}%
Since $\left[ \mathcal{H}_{j},R_{j}\right] =0$, the eigenfunctions could be selected as they have a definite parity, $R_{j}\psi \left(
x_{j}\right) =s_{j}\psi \left( x_{j}\right) $, with $s_{j}=\pm 1$. Then, we set
\begin{equation}
\xi _{j}=m\omega x_{j}^{2};\quad \quad \psi \left( \xi _{j}\right) =\xi
_{j}^{\frac{1-s_{j}}{4}}e^{-\frac{\xi _{j}}{2}}\Upsilon \left( \xi
_{j}\right),
\end{equation}%
thus, Eq. \eqref{B} becomes
\begin{equation}
\left \{ \xi _{j}\frac{\partial ^{2}}{\partial \xi _{j}^{2}}+\left( 1+\mu
_{j}-\frac{s_{j}}{2}-\xi _{j}\right) \frac{\partial }{\partial \xi _{j}}%
+n_{j}\right \} \Upsilon \left( \xi _{j}\right) =0,  \label{pol}
\end{equation}%
where $n_{j}$ is non-negative integer quantum numbers. In this case, the energy eigenvalue function, which depends on parity, is quantized as follows:
\begin{equation}
\mathcal{E}_{n_{j}}^{s_{j}}=2m\omega \left[ 2n_{j}+\left( \frac{1}{2}+\mu
_{j}\right) \left( 1-s_{j}\right) \right] .
\end{equation}%
Then, we can express the general solution of Eq. \eqref{pol} in terms of the associated Laguerre polynomials as
\begin{equation}
\Upsilon \left( \xi _{j}\right) =\mathcal{C}_{s_{j}}\mathbf{L}_{n_{j}}^{\mu
_{j}-\frac{s_{j}}{2}}\left( m\omega x_{j}^{2}\right) ,
\end{equation}
where $\mathcal{C}_{s_{j}}$ is a normalization constant that can be determined from  Eq. \eqref{scalar} as follows \cite{grad}:
\begin{equation}
\int x^{\alpha }e^{-x}L_{n}^{\alpha }\left( x\right) L_{m}^{\alpha }\left(
x\right) =\delta _{nm}\frac{\left( n+\alpha \right) !}{n!}.  \label{I}
\end{equation}%
After a straightforward calculation, we obtain the following expression for
the normalization constant
\begin{equation}
\mathcal{C}_{s_{j}}=\sqrt{\frac{2m\omega n_{j}!}{\left( n_{j}+\mu _{j}-\frac{%
s_{j}}{2}\right) !}}.
\end{equation}%
To evaluate the total expression of the energy eigenvalues, we replace $\mathcal{E}_{n_{j}}^{s_{j}}$ by their expressions defined in Eq. \eqref{E}. We find
\begin{equation}
E_{n}^{\left( s_{1},s_{2},s_{3}\right) }=\pm \sqrt{2m\omega \left[ 2n+\left(
\mu _{1}+\frac{1}{2}\right) \left( 1-s_{1}\right) +\left( \mu _{2}+\frac{1}{2%
}\right) \left( 1-s_{2}\right) +\left( \mu _{3}+\frac{1}{2}\right) \left(
1-s_{3}\right) \right] +m^{2}},  \label{T}
\end{equation}
where $n=n_{1}+n_{2}+n_{3}$. 

Our findings show that the total energy eigenvalue function of the Dunkl-Klein-Gordon oscillator depends not only on the quantum numbers but on the Wigner parameters and parities. The maximal contribution from the additional terms is obtained for $s_{1}=s_{2}=s_{3}=-1$, while the minimal contribution
is achieved for $s_{1}=s_{2}=s_{3}=+1$. 

\begin{figure*}
\resizebox{\linewidth}{!}{\includegraphics{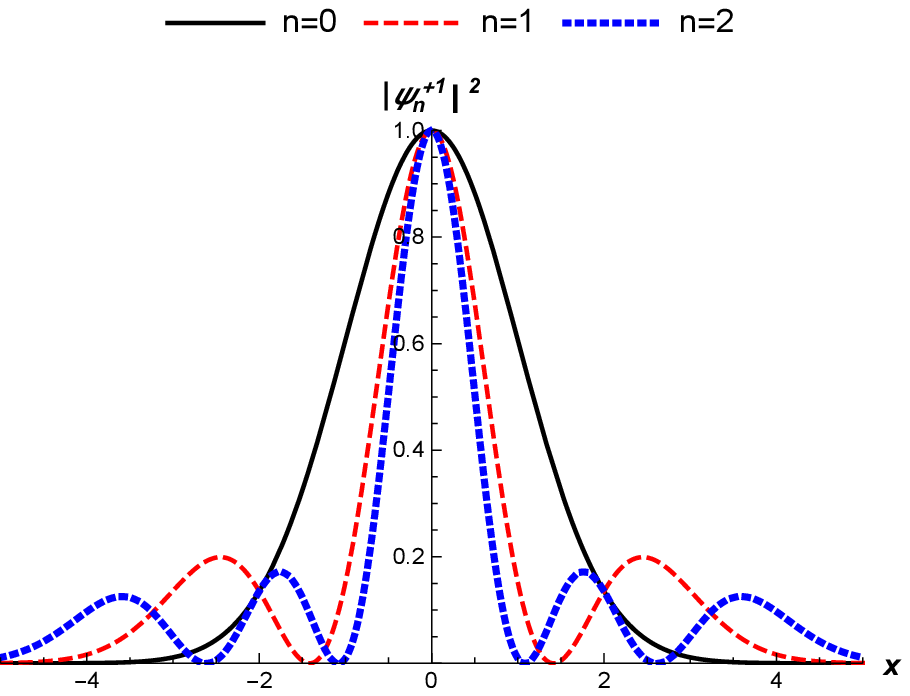},\includegraphics{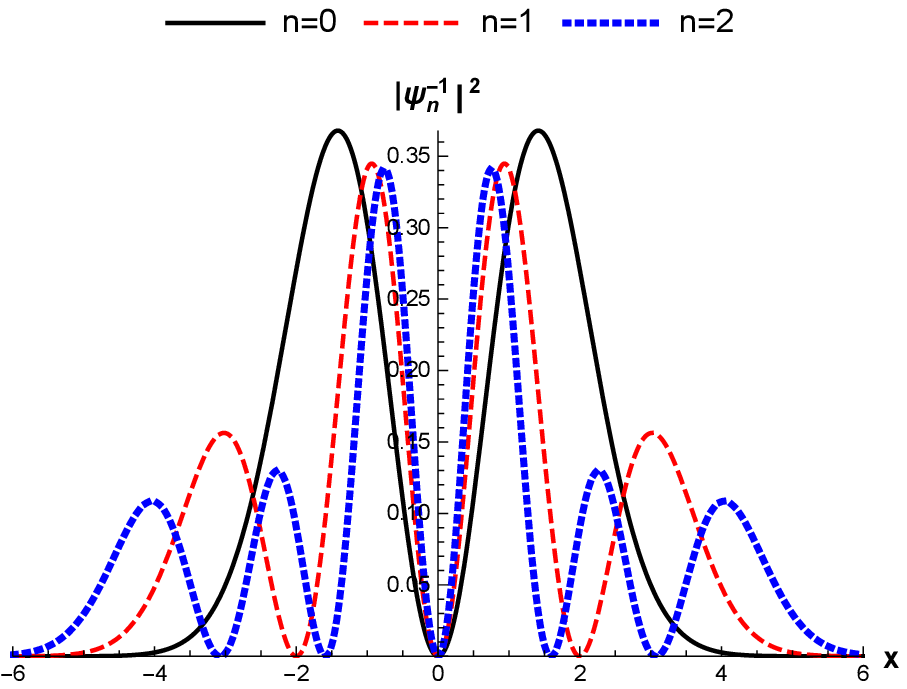}}
\caption{Reduced probability densities versus coordinate.}\label{fig:Fig1}
\end{figure*}

Before we end this section, we present the plots of probability densities of the ground and first two excited states in Fig. \ref{fig:Fig1}. Here, we take $m=0.5$, $\omega=1$ and $\mu_1=0.5$. We demonstrate $s=1$ and $s=-1$ on the right and left columns, respectively. 
We observe that for the positive parity case, the highest probability of detecting particle detection is around the center. Contrary,  in the negative parity case, the highest probability of detection particle on the center is zero.    

\subsection{Solution in Spherical coordinates}
In spherical coordinates, we adopt
\begin{equation}
x_{1}=r\sin \theta \cos \varphi;\quad \quad x_{2}=r\sin \theta \sin \varphi; \quad \quad x_{3}=r\cos \theta .
\end{equation}%
Therefore, Eq. \eqref{A} takes the form%
\small
\begin{equation}
\left \{ \frac{\partial ^{2}}{\partial r^{2}}+\frac{2\left( 1+\mu _{1}+\mu
_{2}+\mu _{3}\right) }{r}\frac{\partial }{\partial r}-m^{2}\omega
^{2}r^{2}+2m\omega \left( \frac{3}{2}+\mu _{1}R_{1}+\mu _{2}R_{2}+\mu
_{3}R_{3}\right) +\frac{\mathcal{J}_{\varphi }}{r^{2}\sin ^{2}\theta }+\frac{%
\mathcal{J}_{\theta }}{r^{2}}+E^{2}-m^{2}\right \} \psi =0,  \label{R}
\end{equation}%
\normalsize
where $\mathcal{J}_{\varphi }$ and $\mathcal{J}_{\theta }$ are
\begin{eqnarray}
\mathcal{J}_{\varphi }&=&\frac{\partial ^{2}}{\partial \varphi ^{2}}+2\bigg[
\mu _{2}\cot \varphi -\mu _{1}\tan \varphi \bigg] \frac{\partial }{\partial
\varphi }-\frac{\mu _{1}}{\cos ^{2}\varphi }\left( 1-R_{1}\right) -\frac{\mu
_{2}}{\sin ^{2}\varphi }\left( 1-R_{2}\right) , \\
\mathcal{J}_{\theta }&=&\frac{\partial ^{2}}{\partial \theta ^{2}}+\cot \theta
\frac{\partial }{\partial \theta }+2\bigg[ \left( \mu _{1}+\mu _{2}\right)
\cot \theta -\mu _{3}\tan \theta \bigg] \frac{\partial }{\partial \theta }-%
\frac{\mu _{3}}{\cos ^{2}\theta }\left( 1-R_{3}\right) .
\end{eqnarray}
At this point we must note that in spherical coordinates the reflection operators act on the wave function as follows:
\begin{equation}
R_{1}\psi \left( r,\theta ,\varphi \right) =\psi \left( r,\theta ,\pi
-\varphi \right) ;\text{ \ }R_{2}\psi \left( r,\theta ,\varphi \right) =\psi
\left( r,\theta ,-\varphi \right) ;\text{ \ }R_{3}\psi \left( r,\theta
,\varphi \right) =\psi \left( r,\pi -\theta ,\varphi \right) .
\end{equation}%
Then, we assume $\psi =\digamma \left(r\right) \Theta \left( \theta \right) \Phi \left( \varphi \right) $ and substitute it in Eq. \eqref{R} to separate it to three ordinary differential equations.  We find
\small
\begin{eqnarray}
\left \{ \frac{\partial ^{2}}{\partial \varphi ^{2}}+2\left( \mu _{2}\cot
\varphi -\mu _{1}\tan \varphi \right) \frac{\partial }{\partial \varphi }-%
\frac{\mu _{1}}{\cos ^{2}\varphi }\left( 1-R_{1}\right) -\frac{\mu _{2}}{%
\sin ^{2}\varphi }\left( 1-R_{2}\right) +\Omega ^{2}\right \} \Phi \left(
\varphi \right) &=&0,  \label{S1} \\
\left \{ \frac{\partial ^{2}}{\partial \theta ^{2}}+\cot \theta \frac{%
\partial }{\partial \theta }+2\left( \left( \mu _{1}+\mu _{2}\right) \cot
\theta -\mu _{3}\tan \theta \right) \frac{\partial }{\partial \theta }-\frac{%
\mu _{3}}{\cos ^{2}\theta }\left( 1-R_{3}\right) -\frac{\Omega ^{2}}{\sin
^{2}\theta }+\varpi ^{2}\right \} \Theta \left( \theta \right) &=&0,  \label{S2} \\
\left \{ \frac{\partial ^{2}}{\partial r^{2}}+\frac{2\left( 1+\mu _{1}+\mu
_{2}+\mu _{3}\right) }{r}\frac{\partial }{\partial r}-m^{2}\omega
^{2}r^{2}+2m\omega \left( \frac{3}{2}+\mu _{1}R_{1}+\mu _{2}R_{2}+\mu
_{3}R_{3}\right) -\frac{\varpi ^{2}}{r^{2}}+E^{2}-m^{2}\right \} \digamma
\left( r\right) &=&0,  \label{S3}
\end{eqnarray}%
\normalsize
where $\Omega ^{2}$, and $\varpi ^{2}$ are two separation constants. We observe that Eqs. \eqref{S1} and \eqref{S2} are similar to the ones which appeared in the study of the three dimensional Dunkl-harmonic oscillator problem \cite{Genest3}. Therefore, we give their solutions in terms of the Jacobi polynomials that are characterized with parity quantum numbers, $s_{1}$, $s_{2}$, $s_{3}$, and their reflection operator correspondences $k=\frac{1-s_{1}}{2}$, $p=\frac{1-s_{2}}{2}$, and $\sigma =\frac{1-s_{3}}{2}$. 
\begin{eqnarray}
\Phi ^{s_{1},s_{2}}\left( \varphi \right) &=& \mathcal{C}_{\varphi }\cos
^{k}\left( \varphi \right) \sin ^{p}\left( \varphi \right) \mathbf{P}_{\nu -%
\frac{k+p}{2}}^{\left( \mu _{2}+p-\frac{1}{2},\mu _{1}+k-\frac{1}{2}\right)
}\left( \cos \left( 2\varphi \right) \right) ,  \label{phi} \\
\Theta ^{s_{3}}\left( \theta \right) &=& \mathcal{C}_{\theta }\cos ^{\sigma
}\left( \theta \right) \sin ^{2\nu }\left( \theta \right) \mathbf{P}_{\ell -%
\frac{\sigma }{2}}^{\left( 2\nu +\mu _{1}+\mu _{2},\mu _{3}+\sigma -\frac{1}{%
2}\right) }\left( \cos \left( 2\theta \right) \right) .  \label{theta}
\end{eqnarray}%
Here $\mathcal{C}_{\varphi }$ and $\mathcal{C}_{\theta }$ are the normalization constants. When $s_{1}.s_{2}=1,$ $\nu $ is a positive integer whereas it is a positive half-integer when $s_{1}.s_{2}=-1$. In a special case $\nu =0$ only if  $s_{1}=s_{2}=1$. Similarly if $s_{3}=1,$ then the quantum number $\ell $ takes only positive integer values whereas they have non-negative half integer values when $s_{3}=-1$. For the validity of the given solutions, the separation constants have to satisfy the following conditions:
\begin{eqnarray}
\Omega^{2}&=&4\nu \left( \nu +\mu _{1}+\mu _{2}\right), \label{cons1} \\
\varpi^{2}&=&4\left( \ell +\nu \right) \left( \ell +\nu +\mu _{1}+\mu _{2}+\mu _{3}+\frac{1}{2}\right) .  \label{cons2}
\end{eqnarray}%
Next, we look for an exact solution to the radial equation. To this end, we first substitute Eqs.\eqref{cons1} and \eqref{cons2} into Eq.(\ref{S3}). We arrive at
\small
\begin{eqnarray}
\bigg\{
\frac{\partial ^{2}}{\partial r^{2}}+\frac{2\left( 1+\mu _{1}+\mu _{2}+\mu
_{3}\right) }{r}\frac{\partial }{\partial r}&-&m^{2}\omega ^{2}r^{2} + 2m\omega
\left( \frac{3}{2}+\mu _{1}s_{1}+\mu _{2}s_{2}+\mu _{3}s_{3}\right)  
 \nonumber \\ &-& \frac{4\left( \ell +\nu \right) \left( \ell +\nu +\mu _{1}+\mu _{2}+\mu_{3}+\frac{1}{2}\right) }{r^{2}}+ (E^{2}-m^{2})\bigg\} \digamma \left( r\right) =0.  \label{r}
\end{eqnarray}%
\normalsize
With the aid of the new variable definition%
\begin{equation}
\rho =m\omega r^{2};\quad \quad \digamma =\rho ^{\nu +\ell }e^{-\frac{\rho }{2}}\Xi \left( \rho \right) ,
\end{equation}%
Eq. \eqref{r} takes the form%
\begin{equation}
\left \{ \rho \frac{d^{2}}{d\rho ^{2}}+\left( \frac{3}{2}+2\left( \nu +\ell
\right) +\mu _{1}+\mu _{2}+\mu _{3}-\rho \right) \frac{d}{d\rho }+\left( \mu
_{1}k+\mu _{2}p+\mu _{3}\sigma +\left( \nu +\ell \right) \right) +\frac{%
E^{2}-m^{2}}{4m\omega }\right \} \Xi =0.
\end{equation}
The solution of the above equation can be given in terms of the Laguerre polynomials as follows%
\begin{equation}
\Xi \left( \rho \right) =\mathcal{C}_{r}\mathbf{L}_{N}^{2\left( \nu +\ell
\right) +\mu _{1}+\mu _{2}+\mu _{3}+\frac{1}{2}}\left( \rho \right) ,\text{ }
\end{equation}%
where the quantum number, $N$, has the form
\begin{equation}
N=-\frac{1}{2}\left[ \mu _{1}\left( 1-s_{1}\right) +\mu _{2}\left(
1-s_{2}\right) +\mu _{3}\left( 1-s_{3}\right) +2\left( \nu +\ell \right) %
\right] +\frac{E^{2}-m^{2}}{4m\omega }. \label{N}
\end{equation}%
Thus, the radial eigenfunctions of the three dimensional Dunkl-Klein-Gordon
oscillator becomes%
\begin{equation}
\text{\ }\digamma \left( \rho \right) =\mathcal{C}_{r}\rho ^{\nu +\ell }e^{-%
\frac{\rho }{2}}\mathbf{L}_{N}^{2\left( \nu +\ell \right) +\mu _{1}+\mu
_{2}+\mu _{3}+\frac{1}{2}}\left( \rho \right),
\end{equation}
where the energy spectrum reads%
\begin{equation}
E_{N,\nu ,\ell }^{s_{1},s_{2},s_{3}}=\pm \sqrt{2m\omega \left[ 2\left( N+\nu
+\ell \right) +\mu _{1}\left( 1-s_{1}\right) +\mu _{2}\left( 1-s_{2}\right)
+\mu _{3}\left( 1-s_{3}\right) \right] +m^{2}}.  \label{SP}
\end{equation}
We observe that the energy spectrum explicitly depends not only on the quantum
numbers ($N,\nu ,\ell $) but the other parameters, ($\mu _{j}$,$s_{j}$), which characterize the Dunkl derivative. Therefore, we conclude that the energy spectrum is dependent on a term originating from the conventional Klein-Gordon oscillator and an additional term originating from the Dunkl derivative. It is worth noting that, as in the previous section, the maximal contribution of the Dunkl term is obtained for $s_{1}=s_{2}=s_{3}=-1$, while the minimal contribution is achieved from $%
s_{1}=s_{2}=s_{3}=+1$. In addition such a correction term, which depends
explicitly on $s_{j}$, lifts the degeneracy of energy levels.

Before we conclude this section, we briefly would like to introduce the orthogonality relation of the angular and radial parts of the wavefunction. Using the following orthogonality relation (\ref{or})
\begin{equation}
\int \psi _{N,\nu ,\ell }^{\left( s_{1},s_{2},s_{3}\right) }\psi _{N^{\prime
},\nu ^{\prime },\ell ^{\prime }}^{\left( s_{1}^{\prime },s_{2}^{\prime
},s_{3}^{\prime }\right) }\left \vert r\sin \theta \cos \varphi \right \vert
^{2\mu _{1}}\left \vert r\sin \theta \sin \varphi \right \vert ^{2\mu
_{2}}\left \vert r\cos \theta \right \vert ^{2\mu _{3}}r^{2}dr\sin \theta
d\theta d\varphi =\delta _{N,N^{\prime }}\delta _{\nu ,\nu ^{\prime }}\delta
_{\ell ,\ell ^{\prime }}\delta _{s_{1},s_{1}^{\prime }}\delta
_{s_{2},s_{2}^{\prime }}\delta _{s_{3},s_{3}^{\prime }}.  \label{or}
\end{equation}
and the integral property given in \eqref
{I} with the following property of the Jacobi polynomials \cite{grad}
\begin{equation}
\int \left( 1-x\right) ^{\alpha }\left( 1+x\right) ^{\beta }P_{N}^{\left(
\alpha ,\beta \right) }\left( x\right) P_{N}^{\left( \alpha ,\beta \right)
}\left( x\right) dx=\frac{2^{1+\alpha +\beta }\Gamma \left( 1+\alpha
+N\right) \Gamma \left( 1+N+\beta \right) }{N!\left( 1+\alpha +\beta
+2N\right) \Gamma \left( 1+N+\alpha +\beta \right) },
\end{equation}%
we find the normalization constants as follows:
\begin{eqnarray}
\mathcal{C}_{\varphi }&=&\sqrt{\frac{4\left( \nu -\frac{k+p}{2}\right) !\left(
1+2\nu +\mu _{2}+\mu _{1}-1\right) \Gamma \left( 1+\nu +\frac{k+p}{2}+\mu
_{1}+\mu _{2}-1\right) }{\Gamma \left( 1+\nu +\mu _{2}+\frac{p-k}{2}-\frac{1%
}{2}\right) \Gamma \left( 1+\nu +\frac{k-p}{2}+\mu _{1}-\frac{1}{2}\right) }}%
, \\
\mathcal{C}_{\theta }&=&\sqrt{\frac{2\left( \ell -\frac{\sigma }{2}\right)
!\left( 1+2\ell +2\nu +\mu _{1}+\mu _{2}+\mu _{3}-\frac{1}{2}\right) \Gamma
\left( 1+\ell +2\nu +\mu _{1}+\mu _{2}+\mu _{3}+\frac{\sigma -1}{2}\right) }{%
\Gamma \left( 1+\ell +2\nu +\mu _{1}+\mu _{2}-\frac{\sigma }{2}\right)
\Gamma \left( 1+\ell +\mu _{3}+\frac{\sigma -1}{2}\right) }},\\
\mathcal{C}_{r}&=&\sqrt{\frac{2m\omega n!}{\left( n+2\nu +2\ell +\mu _{1}+\mu
_{2}+\mu _{3}+\frac{1}{2}\right) !}}.
\end{eqnarray}

%



\section{Coulomb potential} \label{sec:4}

In this section, we consider the Coulomb potential $V\left( r\right) =-
\frac{Ze^{2}}{r}$ and intend to derive a solution in three dimensional Dunkl-Klein-Gordon equation. We start by expressing the stationary Dunkl-Klein-Gordon equation according to \cite{BCL}
\begin{equation}
\left \{ \left( E+\frac{Ze^{2}}{r}\right) ^{2}+\frac{\partial ^{2}}{\partial
r^{2}}+\frac{2\left( 1+\mu _{1}+\mu _{2}+\mu _{3}\right) }{r}\frac{\partial
}{\partial r}+\frac{\mathcal{J}_{\varphi }}{r^{2}\sin ^{2}\theta }+\frac{%
\mathcal{J}_{\theta }}{r^{2}}-m^{2}\right \} \psi =0
\end{equation}%
Then,  we define $\varrho =2\varsigma r=2\sqrt{m^{2}-E^{2}}r$, and use the  angular operator's $\left( \mathcal{J}_{\varphi },\mathcal{J%
}_{\theta }\right)$ eigenvalues, $\left( \Omega ^{2},\varpi ^{2}\right) $, we get
the Dunkl-Klein-Gordon radial equation in the following form
\begin{equation}
\left \{ \frac{d^{2}}{d\varrho ^{2}}+\frac{2\left( 1+\mu _{1}+\mu _{2}+\mu
_{3}\right) }{\varrho }\frac{d}{d\varrho }-\frac{\frac{EZe^{2}}{\varsigma }}{%
\varrho }+\frac{Z^{2}e^{4}-4\left( \ell +\nu \right) \left( \ell +\nu +\mu
_{1}+\mu _{2}+\mu _{3}+1/2\right) }{\varrho ^{2}}-\frac{1}{4}\right \} \Psi
\left( r\right) =0.  \label{AA}
\end{equation}%
For the solution we make an ansatz
\begin{equation}
\Psi \left( \varrho \right) =e^{-\frac{\varrho }{2}}\varrho ^{\eta }\chi
\left( \varrho \right) , \label{ansatz}
\end{equation}%
where the constant $\eta $ has to be determined. To this end, we substitute Eq. \eqref{ansatz} into  Eq. \eqref{AA}. We find
\begin{eqnarray}
\bigg \{
\frac{d^{2}}{d\varrho ^{2}}&+&\left( \frac{2\left( 1+\eta +\mu _{1}+\mu
_{2}+\mu _{3}\right) }{\varrho }-1\right) \frac{d}{d\varrho }-\frac{\left(
1+\eta +\mu _{1}+\mu _{2}+\mu _{3}\right) +\frac{EZe^{2}}{\varsigma }}{%
\varrho } \nonumber 
\\&+&
\frac{\eta \left( \eta -1\right) +2\eta \left( 1+\mu _{1}+\mu _{2}+\mu
_{3}\right) +Z^{2}e^{4}-4\left( \ell +\nu \right) \left( \ell +\nu +\mu
_{1}+\mu _{2}+\mu _{3}+1/2\right) }{\varrho ^{2}}%
\bigg \} \chi \left( r\right) =0.
\end{eqnarray}%
Then, we cancel the term proportional to $\frac{1}{\varrho ^{2}}$. This gives a second order equation for $\eta$. 
\begin{equation}
\eta \left( \eta -1\right) +2\eta \left( 1+\mu \right) +Z^{2}e^{4}-4\left(
\ell +\nu \right) \left( \ell +\nu +\mu +1/2\right) =0.
\end{equation}%
Through a straightforward calculation, we easily determine $\eta$:
\begin{equation}
\eta =-\mu _{1}-\mu _{2}-\mu _{3}-\frac{1}{2}+\sqrt{\left( \mu _{1}+\mu
_{2}+\mu _{3}+2\nu +2\ell +\frac{1}{2}\right) ^{2}-Z^{2}e^{4}}.
\end{equation}
Therefore, the radial equation reduced to the confluent type hypergeometric equation
\begin{equation}
\Bigg \{ \varrho \frac{d^{2}}{d\varrho ^{2}}+\bigg[ 1+\Big( 1+2\big( \eta
+\mu _{1}+\mu _{2}+\mu _{3}\big) \Big) -\varrho \bigg] \frac{d}{%
d\varrho }-\frac{EZe^{2}}{\varsigma }-\Big( 1+\eta +\mu _{1}+\mu _{2}+\mu
_{3}\Big) \Bigg \} \chi =0,  \label{E4}
\end{equation}%
which has solution in terms of the  generalized  Laguerre  polynomials, $\mathbf{L}_{n}^{1+2\left( \eta +\mu _{1}+\mu
_{2}+\mu _{3}\right) }\left( \varrho \right)$, where $n$ is a positive-integer number.
\begin{equation}
n=-\frac{EZe^{2}}{\varsigma }-\left( 1+\eta +\mu _{1}+\mu _{2}+\mu
_{3}\right) .
\end{equation}%
Accordingly we express the energy spectrum of the Dunkl-Klein-Gordon equation with Coulomb potential energy as follows:
\begin{equation}
E_{n,\ell ,\nu }(Z)=m\left\{1+\frac{Z^{2}e^{4}}{\left( n+\frac{1}{2}+%
\sqrt{\left( \mu _{1}+\mu _{2}+\mu _{3}+2\nu +2\ell +\frac{1}{2}\right)
^{2}-Z^{2}e^{4}}\right) ^{2}}\right\}^{-1/2}.  \label{EC}
\end{equation}
We find that the energy spectrum is quantized in terms on the quantum numbers $n,\ell $, $\nu $ and the Wigner parameter $\mu _{j}$ which characterizes the Dunkl algebra. Moreover we derive a constraint,
\begin{eqnarray}
\left( \mu _{1}+\mu _{2}+\mu _{3}+2\nu +2\ell +\frac{1}{2}%
\right) >Ze^{2},
\end{eqnarray}
which is necessary for the existence of physical energy eigenvalues. 

Finally, we explore the Dunkl-fine structure energy. To this end, we Taylor expand Eq. \eqref{EC} in powers of $Z^{2}e^{4}$. We find
\begin{eqnarray}
E_{n,\ell ,\nu }(Z) &\simeq &\allowbreak m\left \{ 1-\frac{1}{2}\frac{Z^{2}e^{4}%
}{\left( n+\mu _{1}+\mu _{2}+\mu _{3}+2\nu +2\ell +1\right) ^{2}}\right.
\notag \\
&&\left. -\frac{\left( Z^{2}e^{4}\right) ^{2}}{\left( n+\mu
_{1}+\mu _{2}+\mu _{3}+2\nu +2\ell +1\right) ^{4}}\left( \frac{1}{2(\mu
_{1}+\mu _{2}+\mu _{3}+2\nu +2\ell) +1}-\frac{3}{8}\right) \right \} . \label{DFS}
\end{eqnarray}
Here, the first term corresponds to the rest energy in natural units. 
Dunkl formalism's contribution to the nonrelativistic energy spectrum arises by the second term. We obtain the Dunkl-fine structure correction as the third term in Eq. \eqref{DFS}. It is worth noting that for particular parameter values
\begin{eqnarray}
\mu_1+\mu_2+\mu_3+2(\nu+\ell)= 5/6,
\end{eqnarray}
the Dunkl-fine structure contribution vanishes. 

\section{Conclusion} \label{sec:5}
In this manuscript, we examine the Dunkl-Klein-Gordon oscillator in three-dimensional spatial space. To this end, we obtain the solutions by employing at first the Cartesian, and then, spherical coordinates. In both cases, we observe that the Dunkl-Klein-Gordon oscillator can easily be separated to three one-dimensional equations. After straightforward algebra, we obtain the energy eigenvalue functions in both coordinate systems. We find that the Dunkl operator modifies the energy eigenvalues. We also derive the normalized eigenfunctions in terms of associate(generalized) Laguerre and Jacobi polynomials. Finally, we study the Coulomb potential in the Dunk-Klein Gordon equation. We show how the energy eigenvalue function and Dunkl-fine structure values are changed under the effect of Dunkl deformation.

\end{document}